\documentstyle[11pt,twoside,epsfig,psfig,paspconf]{article}

\newcommand{\msolar}{\mbox{\,${\rm M}_{\odot}$}}

\begin{document}

\title{Three-Dimensional Evolution of the Galactic Fountain}

\author{Miguel A. de Avillez\altaffilmark{1}}
\affil{Department of Mathematics \& Laboratory of Computational Astrophysics, University of \'Evora}
 
\altaffiltext{1}{mavillez@galaxy.lca.uevora.pt}

\begin{abstract}
Gas that escapes from the Galactic disk, rises into the halo, cools and falls back, constitutes a ``Galactic fountain''. Three-dimensional simulations show that such a fountain model reproduces many of the features that have been observed in the Galaxy and other galaxies such as M31 and M33. Here, these results are reported.
\end{abstract}


%
\section{Introduction}
For more than thirty years it has been known that the Galactic halo is populated with HI clouds, whose origins are still not completely understood. The halo clouds were first detected through 21 cm line emission surveys of the Northern Galactic hemisphere carried out by Muller {\rm et~al.} (1963) and were classified, according to their velocity deviations from that of the local standard of rest (LSR), as intermediate velocity couds (IVCs), having velocity deviations between $-$20 and $-$90 km s$^{-1}$; high velocity clouds (HVCs) with velocity deviations between $-$90 and $-$220 km s$^{-1}$; and very high velocity clouds (VHVCs) with velocities above $-$220 km s$^{-1}$ (for a thorough review of these HI clouds, see Wakker (1990) and Wakker \& van Woerden (1997)).

Very little is known about the origin of the halo clouds. This is principally due to a lack of knowledge of their distances. However, some insight regarding their origins has been provided by the combination of 21-cm emission data with absorption line data. These studies have shown that the metallicity of some of
these clouds (although, see Lu et~al. 1998 and Wakker et~al. 1999 for
counter-examples) is similar to that found in disk gas (de Boer {\rm et~al.} 1991), implying that these clouds are not formed from primordial gas, but from gas expelled from the Galactic disk by some energetic events such as supernovae. 

\subsection{Models for the Origin of HI Clouds}
Two distinct classes of models have been proposed for the outflow, depending on the distribution of supernovae in the Galactic disk. The first class, referred to as ``Galactic fountains'', assumes that the flow results from the hot gas, heated by the isolated supernovae in the Galactic disk, forming a continuous outflow  on the Galactic scale (Shapiro \& Field 1976). The second class, referred to as ``Galactic chimneys'', assumes that the flow of gas has its origin in clusters of supernovae that have blown out a hole in the Galactic disk, through which the hot gas flows ballistically into the halo (Tomisaka \& Ikeuchi 1986). 
\subsubsection{Galactic Fountain}
Models of gas flow on the galactic scale were first introduced by
Shapiro \& Field (1976) and subsequently developed by Bregman (1980),
Kahn (1981) and others. The Galactic fountain originates from the widespread supernovae that warm up the disk gas to temperatures of $10^{6}$ K, and flows at a rate of $10^{-19}$ g cm$^{-2}$ s$^{-1}$ into the halo (Bregman 1980; Kahn 1981). The upflowing gas cools and condenses into neutral hydrogen clouds that rain onto the disk, with velocities of the order of 60 to 100 km s$^{-1}$ (Kahn 1981). 
The models assume the height to which the hot gas will rise and the expected rate of condensation in the cooling gas depend only on the temperature of the gas at the base of the fountain and on the rate of cooling of the upflowing gas.

Kahn (1981) showed analytically that most of the gas entering a 
low temperature fountain would be transonic, i.e., the intercloud medium 
would become supersonic at some point not far from the Galactic disk. 
The location of a sonic level for the fountain gas would determine the initial 
conditions of the flow as it enters the fountain, and therefore would be 
the cause for any asymmetry that may exist between hemispheres (see discussion in Avillez {\rm et~al.} 1995). In this model the ascending and descending parts of the fountain gas are separated by a shock wave. The ascending flow entering the shock wave is heated and becomes part of a hot layer that supports the descending cool layer on top of it. Rayleigh-Taylor instabilities grow at the interface separating the two layers. In consequence, the cool gas breaks up into cloudlets falling towards the disk with intermediate velocities.

Houck \& Bregman (1990) using two-dimensional quasi-static models, confirmed the formation of intermediate clouds in a transonic fountain as predicted by Kahn (1981). After the clouds formed they were removed from the computational grid and were treated as independent entities moving in a path governed by gravitational and centrifugal effects. Using this procedure, Bregman (1980) was able to reproduce partially the distribution of HI clouds in the halo. The drawback of these models is the absence of interactions between clouds and the rest of the flow as a result of the clouds being removed from the computational domain.

More sophisticated models considering a two dimensional evolution of the disk and halo gas under the effects of stellar winds and supernova heating have been developed by Rosen {\rm et~al.} (1993) and Rosen \& Bregman (1995). Two co-spatial fluids representing the stars and gas in the interstellar medium were used. These models reproduced the presence of a multiphase media with cool, warm and hot intermixed phases having mean scale heights compatible with those observed in the Galaxy. However, the structure and properties of the ISM changed according to the overall energy injection rate. As a consequence, the models were unable to reproduce, in the same simulation, both the structure of the ISM near the disk and, at the same time, the presence of HI gas with high and intermediate velocities. 
\subsubsection{Localized Outbursts: ``Chimneys''}
HI emission surveys of the Galactic disk by Heiles (1984) and Koo {\rm et~al} (1992) revealed the presence of holes in the emission maps. These holes have sizes varying between a few hundred parsec and a kilo parsec. Surveys of nearby spiral galaxies (M31 and M33) by Brinks \& Bajaja (1986) and Deul \& Hartog (1990) have shown the presence of holes with similar sizes. The holes are associated to isolated as well clustered supernovae running along the spiral arms of the galaxies.

Supernovae in OB associations generate large cavities of low density gas in the Galactic disk. These cavities are surrounded by a thin shell of cold gas swept up by the blast waves of successive supernovae that occur inside the cavity. As a consequence, the shell is accelerated upwards and Rayleigh-Taylor instabilities develop at shell cap leading to its disruption. The hot gas inside the cavity escapes into the halo in a highly energetic outflow, breaking through the Lockman layer. In the disk and lower halo the outflow is confined to a cone-shaped structure by the remains of the old shell (Norman \& Ikeuchi 1989). The energies involved in such a phenomenon are of the order of $10^{53}$ erg (Heiles 1991). As the hot gas rises into the halo, it cools and returns to the Galactic disk, forming a fountain. The height to which the gas rises varies between 5 and 10 kpc. Therefore, chimneys can account for the presence of HI clouds at greater heights than the large scale outflows previously discussed.

Two-dimensional modelling has been carried out by Tomisaka \& Ikeuchi (1986), Mac Low {\rm et~al} (1989) and Tenorio-Tagle {\rm et~al} (1990) using the Dickey-Lockman (1990) gas distribution in the halo. These authors have shown that superbubbles are able to break through the Lockman layer, provided they are generated in OB associations that have a high number of stars and are displaced from the Galactic plane to a height of at least 100 pc. The number of supernovae needed for such
a multiple explosion varies between 50 and 100 (Tenorio-Tagle \& Bodenheimer 1988; Tenorio-Tagle {\rm et~al} 1990). Events like this must be regarded as unusual, although there is evidence for such events in other galaxies (Meaburn 1980).

\subsection{Objectives of the Study}
It is clear that substantial progress has been made in modelling the
Galactic fountain since it was first introduced by Shapiro \& Field in 1976. 
The models described in the previous sections give a two-dimensional description of the evolution of the disk and halo gas, and simulate the vertical distribution of IVCs and HVCs in the halo. However, such two-dimensional calculations impose natural limitations on the evolution of the gas and clouds. The absence of a third dimension constrains the motion to a vertical plane perpendicular to the Galactic disk. If three-dimensional evolution is considered, the overall structure of the flow may suffer modifications that lead to the generation of phenomena that may not be identified in the two-dimensional calculations. Furthermore the models have considered simplified conditions for the stars and ISM. The Galactic fountain models assumed that supernovae were randomly distributed in the disk and occurred with a rate of 3 per century, corresponding to a massflux of hot ($T\sim 10^{6}$ K) material into the halo of the order of $10^{-19}$ g cm s$^{-1}$, whereas global models such as !
those of Rosen \& Bregman (1995) showed that at this rate the models are unable to reproduce the presence of HVCs in the halo. The chimney models neglected the presence of a dynamical thick disk which would provide constraints to the structure of the chimneys. The formation of the chimneys was carried out in a medium where no other phenomena was present. 

The principal objective of this research is to develop a three-dimensional model to account for the collective effects of type Ib, Ic and II supernovae on the structure of the interstellar medium in the galactic disk and halo, and the formation of the major features already observed in the Galaxy; these include the Lockman and Reynolds layers, large scale outflows, chimneys, and HI clouds, features that all contribute to the overall disk-halo-disk cycle known as the Galactic fountain.

In section 2 the numerical modelling used in this study is presented followed by a discussion on the evolution of the simulations in section 3. Section 4 deals with the IVCs and HVCs detected in the simulations. In section 5 a discussion of on the results is carried out, followed by a summary of the fountain model (section 6). 
\section{Numerical Modelling}
\subsection{Model of the Galaxy}
The study of the evolution of the Galactic disk rests in the realization that the Milky Way has a thin and thick disks of gas in addition to a stellar disk. The thin gas disk has a characteristic thickness comparable to that of the stellar disk of Population I stars. The thick gas disk is composed of warm neutral and ionized gases with different scale heights - 500 pc (Lockman {\rm et~al.} 1986) and 950 pc (Reynolds 1987), respectively. 

The stellar disk has a half thickness of 100 pc and the vertical mass distribution, $\rho_{\star}$, inferred from the stars kinematics, given by (Avillez {\rm et~al.} 1997),
\begin{equation}
\label{eq1}
\rho_{\star}=\rho_{\star,\circ} sech ^{2}\left[\left(2\pi G \beta_{\star}\rho_{\star}\right)^{1/2}z\right]
\end{equation}
where $z$ varies between $-100$ pc and $100$ pc, $\rho_{\star,\circ}=3.0\times 10^{-24}$ g cm$^{-3}$ is the mass density contributed by Population I stars near the Galactic plane (Allen 1991) and the constant $\beta_{\star}=1.9\times 10^{-13}$ cm$^{-2}$ s$^{2}$. This mass distribution generates a local gravitational potential, $\Phi$, of the form
\begin{equation}
\label{eq2}
\Phi=-\frac{2}{\beta_{\star}}\ln \cosh\left[\left(2\pi G\rho_{\star}\beta_{\star}\right)^{1/2}z\right].
\end{equation}

The stellar disk is populated by supernovae types Ib, Ic and II, whose major fraction occurs along the spiral arms of the Galaxy close to or in HII regions (Porter \& Filippenko 1987). Supernovae types Ib and Ic have progenitors with masses $M\geq 15\msolar$, whereas Type II SNe originate from early B-type precursors with $9 \msolar\leq M\leq 15 \msolar$. 

The rates of occurrence of supernovae types Ib+Ic and II in the Galaxy are $2\times 10^{-3}$ yr$^{-1}$ and $1.2\times 10^{-2}$ yr$^{-1}$, respectively (Cappellaro {\rm et~al.} 1997). Similar rates have been found by Evans {\rm et~al.} (1989) in a survey of 748 Shapley-Ames galaxies. The total rate of these supernovae in the Galaxy is $1.4\times 10^{-2}$ yr$^{-1}$. $60\%$ of these supernovae occur in OB associations, whereas the remaining $40\%$ are isolated events (Cowie {\rm et~al.} 1979). 

\subsection{Basic Equations and Numerical Methods} 
The evolution of the disk gas is described by the equations of conservation of mass, momentum and energy:
\begin{equation}
\label{eq3}
\frac{\partial \rho}{\partial t}+\nabla\left(\rho {\bf v}\right)=0;
\end{equation}
\begin{equation}
\label{eq4}
\frac{\partial\left( \rho {\bf v}\right)}{\partial t}+{\bf \nabla}\left(\rho {\bf v}{\bf v}\right)=-{\bf \nabla}p-\rho {\bf \nabla}\Phi;
\end{equation}
\begin{equation}
\label{eq5}
\frac{\partial \left(\rho e\right)}{\partial t}+{\bf \nabla}\left(\rho e {\bf v}\right)=-p{\bf \nabla}{\bf v}-n^{2}\Lambda;
\end{equation}
where $\rho$, $p$, $e$, and ${\bf v}$ are the mass density, pressure and specific energy and velocity of the gas, respectively. The set of equations is complete with the equation of state of ideal gases. 

$\Lambda$ is the functional approximation of the cooling functions of Dalgarno \& McCray (1972), Raymond {\rm et~al} (1976) and the isochoric curves of Shapiro \& Moore (1976), except in the range of temperatures of $10^{5}-5\times 10^{6}$ K where the simple power law of the temperature (Kahn 1976)
\begin{equation}
\Lambda=1.3\times 10^{-19} T^{-0.5}\quad\mbox{erg cm$^{3}$ s$^{-1}$}
\end{equation}
is applied. Between $5\times 10^{6}$ and $5\times 10^{7}$ the cooling function has been approximated by $T^{-0.333}$ (Dorfi 1997). In order to avoid any cooling below zero temperature (this is known as catastrophic cooling), $\Lambda$ is zero below 200 K.

The equations of evolution are solved by means of a three-dimensional hydrodynamical scheme using the piecewise parabolic method (Collela \& Woodward 1984) and the adapted mesh refinement algorithm of Berger \& Collela (1989). 

\subsection{Computational Domain and boundary Conditions}
The simulations were carried out using a Cartesian grid centered on the Galactic plane with an area of 1 kpc$^{2}$ and extending from -4 kpc to 4 kpc. Grid resolutions of 5 pc and 10 pc were used for $-270\leq z\leq 270$ pc and for $z\leq -250$ pc and $z\geq +250$ pc, respectively. The cells located at $-270\leq z\leq -250$ pc and $250\leq z\leq 270$ pc of the coarser and finer grids overlap to ensure continuity between the two grids. The solution in the finer grid, at ranges $-270\leq z\leq -250$ pc and $250\leq z\leq 270$ pc, results from a conservative interpolation from the data in the coarser cells to the fine grid cells as prescribed by Berger \& Collela (1989). 

The boundary conditions along the vertical axis are periodic; outflow boundary conditions are used in the upper and lower parts of the grid parallel to the Galactic plane.

\subsection{Initial Conditions}
The initial configuration of the system in the computational domain considers the vertical distribution of the disk and halo gases in accordance to the Dickey \& Lockman (1990) profile, and reproduces the distribution of the thin thick disk gas in the solar neighborhood. The gas is initially in hydrostatic equilibrium with the gravitational field, thus requiring an initial gas temperature as defined by the density distribution.

Throughout the simulations the presence of the stellar disk is required with a thickness of 200 pc and an area of 1 kpc$^{2}$, corresponding to a volume of $V=2\times 10^{8}$ pc$^{3}$. Isolated and clustered supernovae ot types Ib, Ic and II occur in the volume $V$ at rates obtained from the values discussed in \S2.1 and given by   
\begin{equation}
\sigma\frac{V}{V_{G}}=1.42\times 10^{-3}\sigma \mbox{\hspace*{0.2cm} yr$^{-1}$} \end{equation}
where $V_{G}=1.4\times 10^{11}$ pc$^{3}$ is the volume of the Galactic disk with a radius of 15 kpc and a thickness of 200 pc, and $\sigma$ is the rate of supernovae observed in the volume $V_{G}$. Table 1 presents the rates of supernovae in the Galaxy as well as in the simulated disk. In the simulated stellar disk isolated and clustered supernovae form at time intervals of $1.26\times 10^{5}$ yr and $8.4\times 10^{4}$ yr, respectively.
\begin{table}[thbp]
\label{tab1}
\begin{center}
\begin{tabular}{c|c|c|c|c}
\hline
Volume & Distribution & $\sigma_{Ib+Ic}$ & $\sigma_{II}$ & $\large \sigma_{Ib+Ic+II}$\\
 & & \small ($10^{-3}$ yr$^{-1}$) & \small ($10^{-3}$ yr$^{-1}$) & \small ($10^{-3}$ yr$^{-1}$)\\
\hline
\hline
\multicolumn{1}{c|}{Galaxy} & OB Assoc. & 1.2 & 7.2 & 8.4\\ 
\cline{2-5}
			    & Isolated 	& 0.8 & 4.8 & 5.6 \\
\hline
\hline
\multicolumn{1}{c|}{Sim. Disk} & OB Assoc. & 0.002 & 0.01 & 0.012\\ 
\cline{2-5}
& Isolated & 0.001& 0.007 & 0.008\\
\hline
\end{tabular}
\caption{Rates of occurrence of supernovae types Ib+Ic and II in the Galactic disk having a volume $V_{G}$ and in the simulated stellar disk with volume $V$. The rates account fos isolated as well as clustered supernovae.}
\end{center}
\end{table}
The time interval between two successive superbubbles in the volume $V$ varies between $1.9\times 10^{6}$ yr and $5.23\times 10^{6}$ yr (Ferri\`ere 1995), but following Norman and Ikeuchi (1989), and adopting an average value of $N_{SN}=30$ supernovae per superbubble the time interval of formation of superbubbles in the volume $V$ is
\begin{equation}
\frac{N_{SN}}{\sigma_{OB}}=2.5\times 10^{6}\;\mbox{yr},
\end{equation}
the value used in this study.
\subsection{Numerical Setup of Supernovae}
Each supernova is setup in the beginning of phase II, with a thermal energy content of (Kahn 1975)
\begin{equation}
E_{Therm}=0.36\rho_{\circ}a
\end{equation}
and a kinetic energy content of
\begin{equation}
E_{Kin}=0.138\rho_{\circ}a,
\end{equation}
where $a=2E/\rho_{\circ}$, $\rho_{\circ}$ is the local density of the medium where the supernova occurs and $E$ is the energy of the explosion.

The generation of supernovae in the stellar disk is carried out in a semi-random way where the coordinates of the supernova are determined from three new random numbers converted to the dimensions of the stellar disk. The distances are mapped anywhere within the volume of a cell in the grid.

The location of a new supernova in the grid is based in the following constraints: (a) vertical density distribution of population I stars given by equation (\ref{eq1}), (b) type of newer supernova, (c) rates of occurrence of supernovae types Ib+Ic and II and (d) rate of formation of superbubbles. 

When a supernova is due to appear in a specific location, a supernova generator checks for the occurrence of a previous supernova at that location. If this test is negative, then a new supernova is setup. If the test is positive, then the time delay between the old supernova and the new one is checked. Now the decision is based on the comparison of the this time delay and that of a new supernova within a superbubble. If the delay is smaller than the time delay within the superbubble, then no supernova is generated and the generator chooses a new position from a new set of random numbers, else a supernova is setup up.

Using simple rules, the generator is able to form supernovae in the stellar disk with a distribution compatible with observations and, at the same time, defining the type of supernovae and thus the mass loaded into the surrounding medium during its explosion.
\section{Evolution of the Simulations}
The simulations were carried out for a period of 1 Gyr with supernovae being generated at time 0. 
\begin{figure}[thbp]
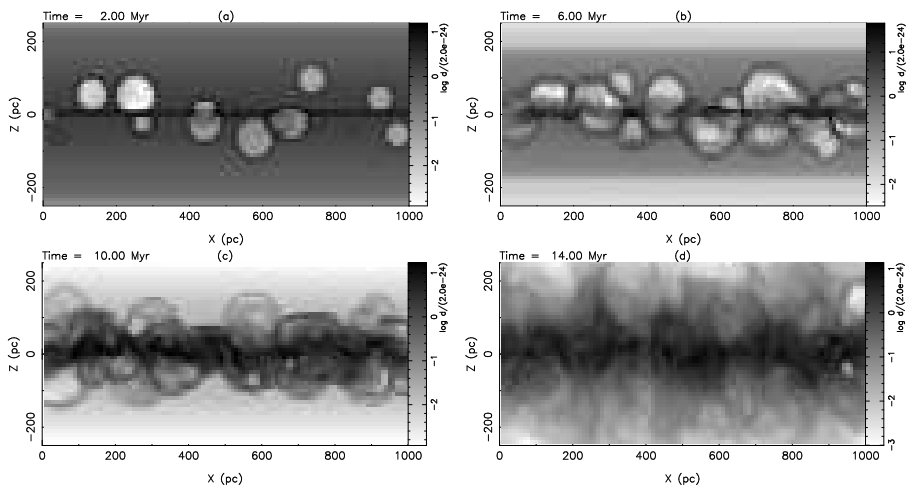

\centering{\hspace*{-0.1cm}}
\psfig{file=mavillez-fig1ab.ps,angle=-90,width=4.7in,clip=}
\psfig{file=mavillez-fig1cd.ps,angle=-90,width=4.7in,clip=}
\caption{\small Grey-scale maps of the density distribution during the first 25 Myr for $y=500$ pc at: (a) 2 Myr, (b) 6 Myr, (c) 10 Myr and (d) 14 Myr after start of calculations. Note the collapsing of the gas during the first 10 Myr. As the cold gas descends towards the plane it leaves behind a low density medium (light grey in images (b) and (c)).} 
\end{figure}
In general the simulations start from a state of hydrostatic equilibrium breaking up in the first stages of the simulation. The system then evolves into a statistical steady equilibrium where the overall structure of the ISM seems similar on the global scale.

Initially there is an imbalance between cooling and heating of the gas in the disk giving rise to an excess of radiative cooling over heating because of the small number of supernovae during this period. The gas originally located in the lower halo starts cooling and moves towards the midplane, colliding there with gas falling from the opposite side of the plane as can be seen in Figure 1. The figure shows grey-scale maps of the density distribution around the Galactic plane taken at $y=500$ pc at four times: 2, 6, 10 and 14 Myr. After 10 Myr of evolution, the major fraction of the initial mass is confined within a slab having a characteristic thickness of 100-150 pc (Figure 1 (c)). As the supernovae occur they warm up the gas in the slab gaining enough energy to overcome the gravitational pull of the disk and therefore, expanding upwards (Figure 1 (d)) redistributing matter and energy in the computational domain.
\begin{figure}[thbp]
\centering
\mbox{\epsfxsize=3.2in\epsfysize=2.5in\epsfbox{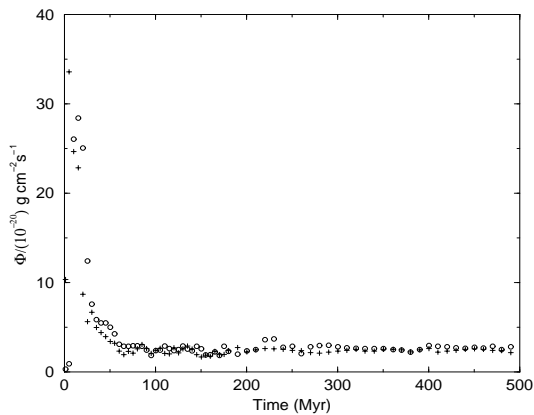}}
\caption{Total massflux of the ascending (circles) and descending (crosses) flows measured at $z=140$ pc and averaged over the entire area of the Galactic disk. A balance between the ascending and descending flows sets in after 50 Myr.}
\end{figure}

Figure 2 presents the total massflux of the descending and ascending gases measured at $z=140$ pc and averaged over the entire area of the Galactic disk during the first 500 Myr of evolution. During the first 10-20 Myr the descending gas has a maximum value of $3.4\times 10^{-19}$ g cm$^{-2}$ s$^{-1}$ (that is $4.8\times 10^{-2}$ $\msolar$ kpc$^{-2}$ yr$^{-1}$). The ascending gas massflux peaks at $2.8\times 10^{-19}$ g cm$^{-2}$ s$^{-1}$ ($3.9\times 10^{-2}$ $\msolar$ kpc$^{-2}$ yr$^{-1}$). The expansion of the disk gas diminishes during the next 20 Myr, with the consequent decrease in the total massflux. At 50 Myr a balance between descending and ascending flows sets in. Both the ascending and descending gases have massfluxes of approximately $3\times 10^{-20}$ g cm$^{-2}$ s$^{-1}$ ($4.2\times 10^{-3}$ $\msolar$ kpc$^{-2}$ yr$^{-1}$), which corresponds to a total inflow rate of 2.97 $\msolar$ yr$^{-1}$ in the Galaxy on one side of the Galactic plane.

This balance is observed during the rest of the simulations indicating that a quasi-equilibrium between the descending and ascending gases dominates the evolution of the disk-halo gas. However, there are some periods when one of the flows dominates over the other. This is related with periodic imbalances between cooling and heating leading to compressions and expansions of the disk gas. Such an effect has already been found in the two-dimensional simulations of the interstellar medium carried out by Rosen \& Bregman (1995).

\subsection{Galactic Disk}
After the first 100 Myr the system evolves in such a way to approach a statistical steady equilibrium where the dynamics of the gases in the disk and the halo are similar throughout the simulations. As such, the ISM will have the same structure at different times and locations in the disk and halo, although it experiences changes due to local processes such as supernovae, collisions with HI clouds, etc. In order to stress this point, pictures showing the time evolution of the disk gas between 200 and 500 Myr are presented (Figures 3-7).
\begin{figure}[thbp]
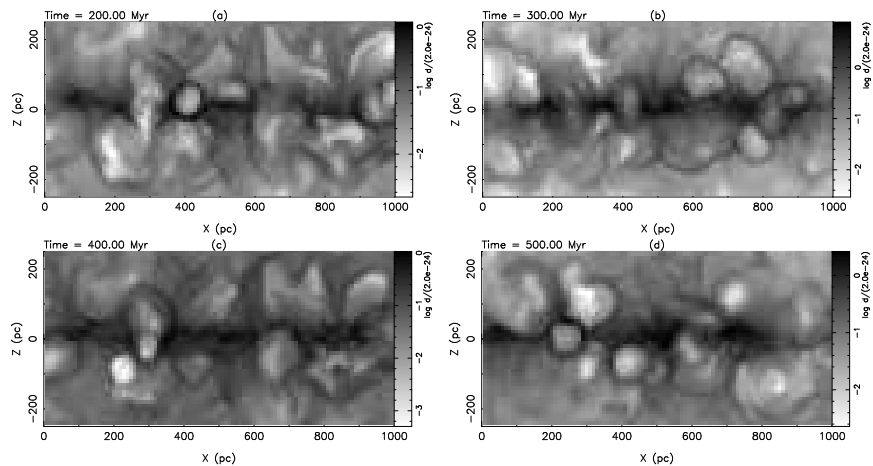

\centering{\hspace*{-0.1cm}}
\psfig{file=mavillez-fig3ab.ps,angle=-90,width=4.5in,clip=}
\psfig{file=mavillez-fig3cd.ps,angle=-90,width=4.5in,clip=}
\caption[Gray scale maps of the gas density for y=30 pc]{Gray scale maps of the gas density for y=30 pc at: (a) 200 Myr, (b) 300 Myr, (c) 400 Myr, and (d) 500 Myr after start of calculations.}
\end{figure}

The images show the distribution of the gas density at different locations in the disk. Edge-on maps of the disk gas for $\left|z\right|\leq 250$ pc, taken at $y=30$ pc and $y=790$ pc (Figures 3 and 4) show a disk composed of a multiphase medium where cool ($T< 8000$ K), warm ($8000\leq T\leq 10^{5}$ K) and hot ($T> 10^{5}$ K) phases co-exist. The major fraction of the disk is filled with warm and hot gas. Most of that gas originates from supernovae spread over the disk. Embedded in this medium are cold sheets and clouds with various sizes and forms. The cold gas is mainly confined to the thin disk with a thickness varying between 20 and 50 pc and having a wiggly structure resulting from supernovae or collisions of descending clouds with the disk gas.
\begin{figure}[thbp]
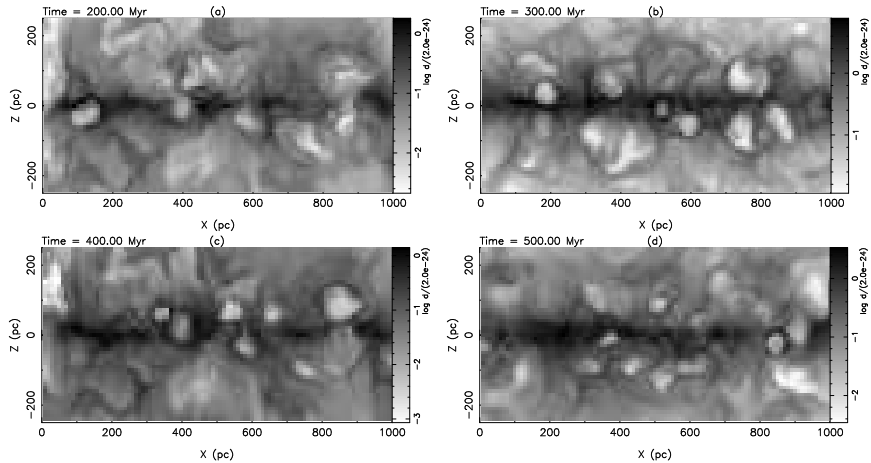

\centering{\hspace*{-0.1cm}}
\psfig{file=mavillez-fig4ab.ps,angle=-90,width=4.5in,clip=}
\psfig{file=mavillez-fig4cd.ps,angle=-90,width=4.5in,clip=}
\caption[Gray scale maps of the gas density for y=790 pc]{Gray scale maps of the gas density for y=790 pc at: (a) 200 Myr, (b) 300 Myr, (c) 400 Myr, and (d) 500 Myr after start of calculations.}
\end{figure}
Sheets wiggling in directions perpendicular to the plane, resembling ``worms'' crawling out of the disk, are observed in all the images. They are associated with broken shells or supershells surrounding cavities of low density gas ($\rho\sim 10^{-26}$ g cm$^{-3}$) generated by isolated and correlated supernovae, respectively.

Clouds result from broken shells or from cool gas that, during its descent towards the disk, is compressed by the interaction with the hot gas flowing upwards in the disk. The interaction of shock waves increase the local density by factors of four times the gas ahead the shock. Therefore, triggering the formation of clouds and sheets.  

Face-on maps of the disk gas taken at $z=\pm250$ pc (Figures 5 and 6) show the presence of clouds as well as sheets with thicknesses of 5 pc and widths of several tens of parsec even hundreds of parsec. Due to the resolution of the calculations in the disk clouds with dimensions smaller than 5 pc are not resolved in the images.
\begin{figure*}[thbp]
\centering{\hspace*{-0.1cm}}
\psfig{file=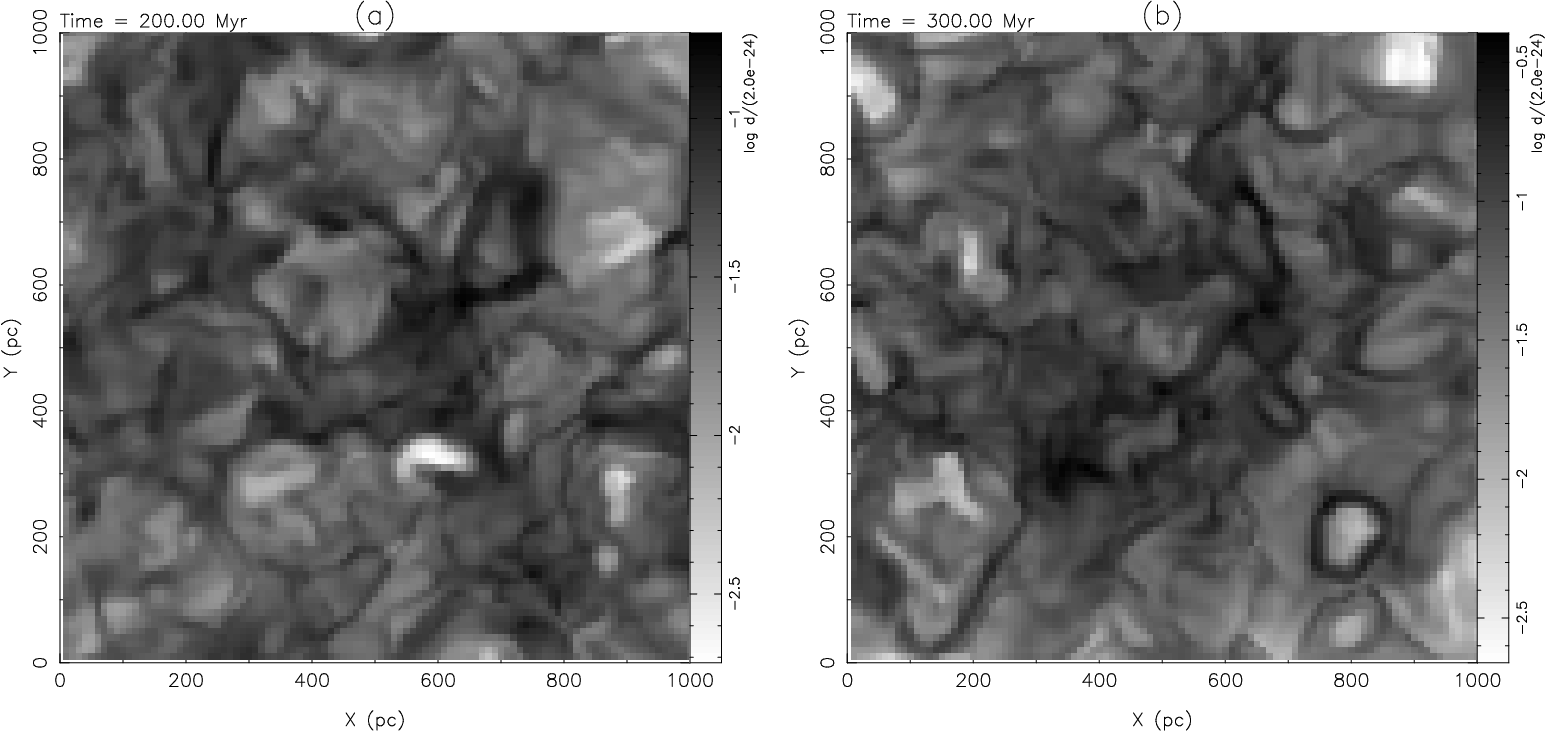,angle=-90,width=4.5in,clip=}
\caption{Gray scale maps of the gas density for z=-250 pc at: (a) 200 Myr and (b) 300 Myr after start of calculations.}
\centering{\hspace*{-0.1cm}}
\psfig{file=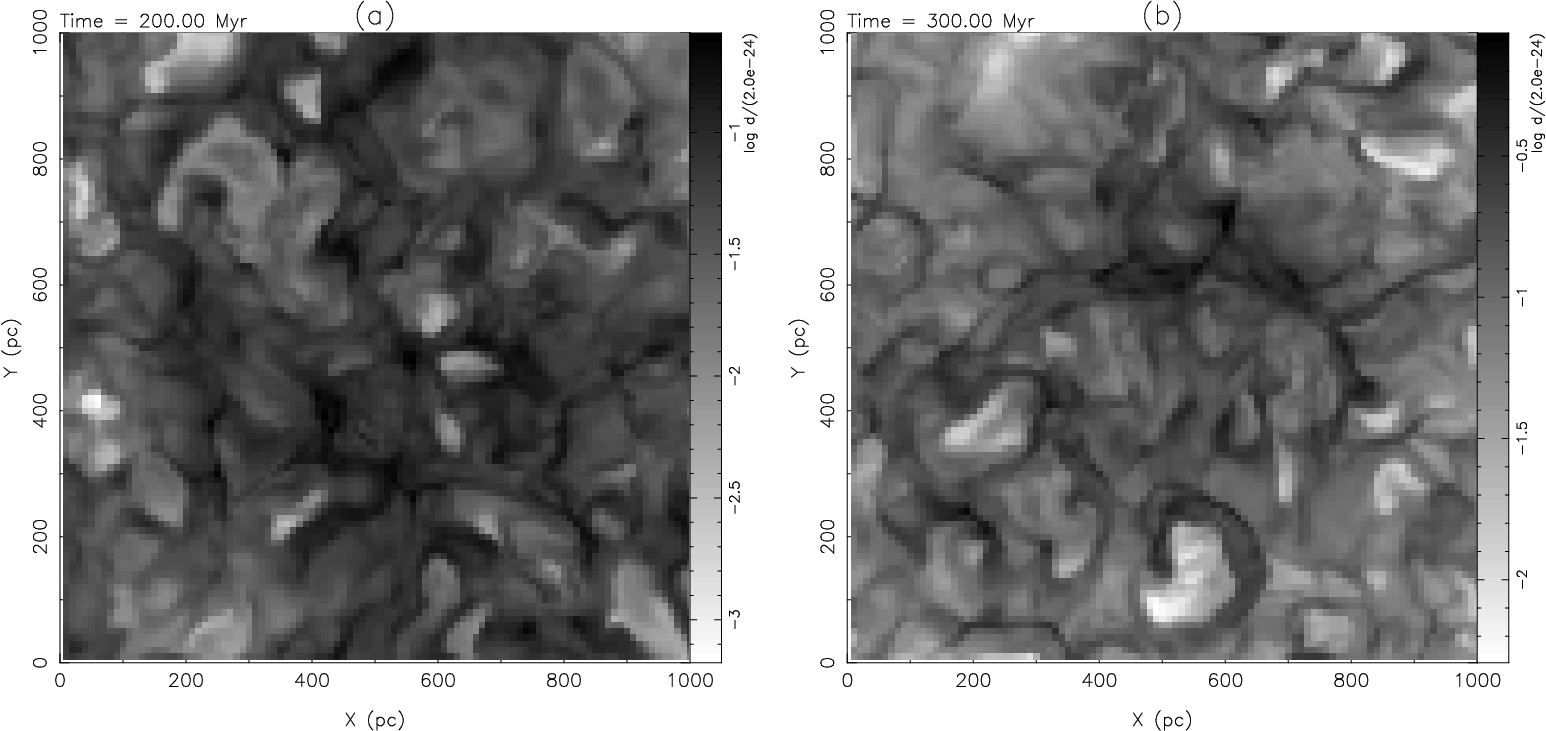,angle=-90,width=4.5in,clip=}
\caption{Gray scale maps of the gas density for z=250 pc at: (a) 200 Myr and (b) 300 Myr after start of calculations.}
\centering{\hspace*{-0.1cm}}
\psfig{file=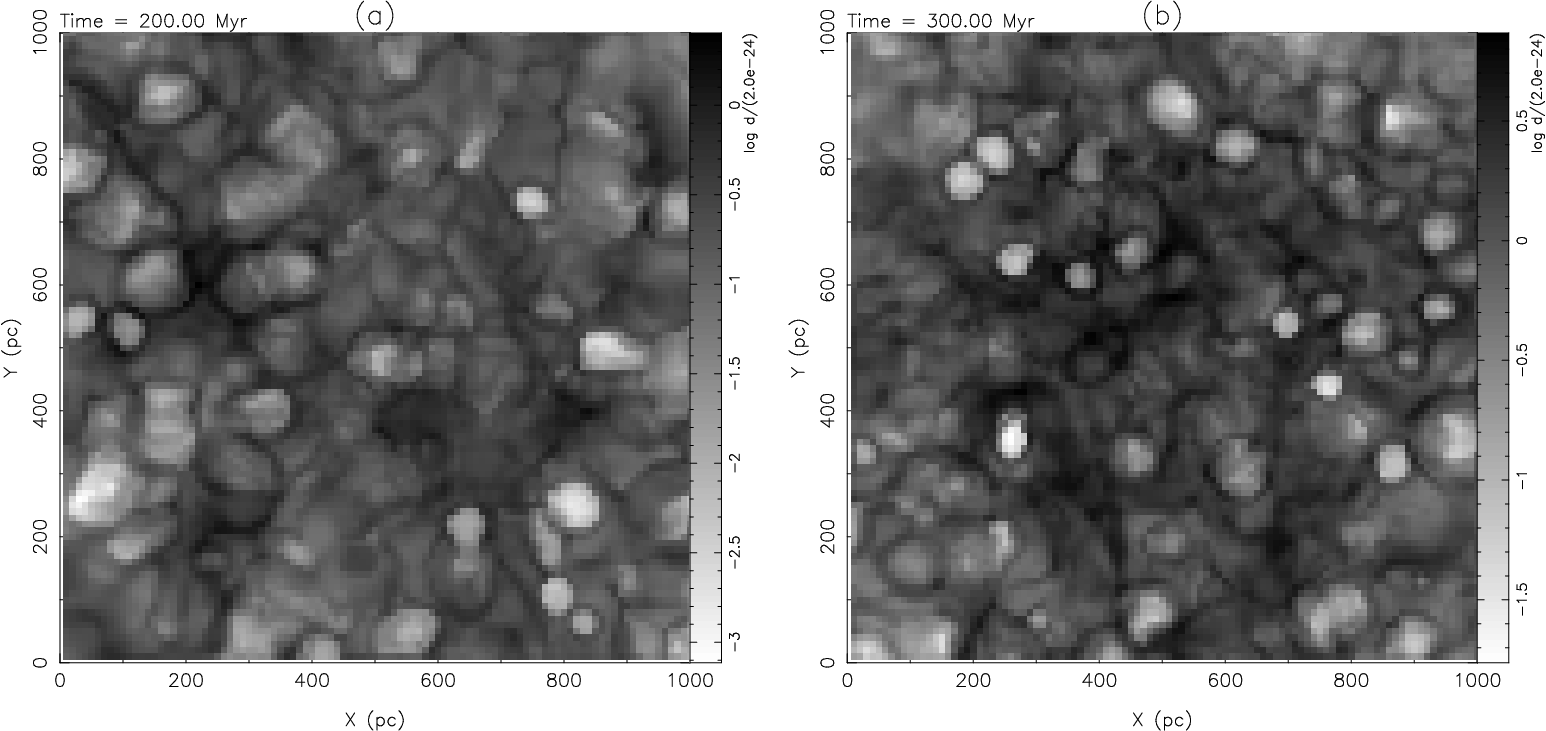,angle=-90,width=4.5in,clip=}
\caption[Gray scale maps of the gas density for $z=0$ pc]{Gray scale maps of the gas density for z=0 pc at: (a) 200 Myr and (b) 300 Myr after start of calculations.}
\end{figure*}
The distribution and sizes of the sheets varies with the increase of $z$ within the disk. In the midplane the sheets are thin and short as result of the interaction of supernovae and collision with other clouds (Figure 7). 

As the blast waves from supernovae expand in the disk, they swept up clouds and sheets triggering their disruption into smaller structures. At $z=\pm 250$ pc the presence of sheets results from: (a) cold gas descending from above and having sheet-like forms acquired during their formation as result of Rayleigh-Taylor instabilities that occurred in larger clouds and (b) breaking up of shells and supershells that expanded upwards and are displaced from the Galactic plane.
\begin{figure*}[thbp]
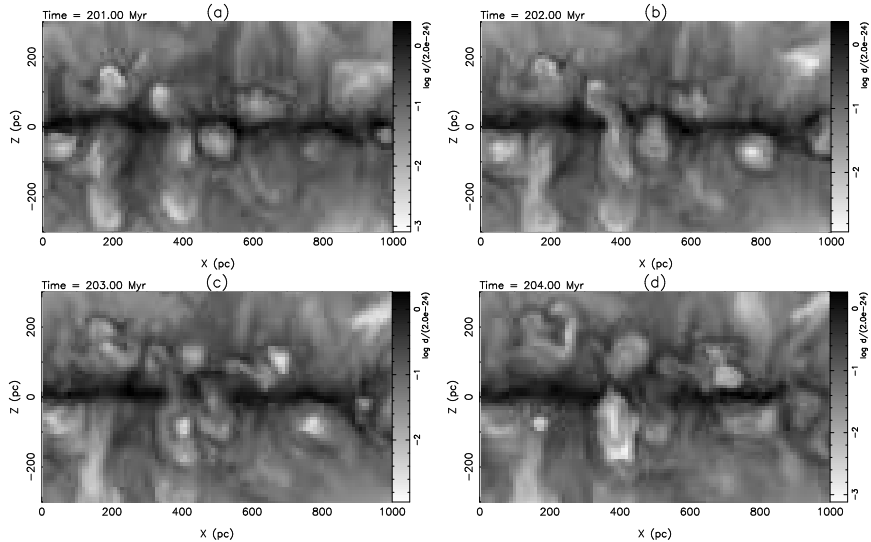

\centering{\hspace*{-0.1cm}}
\psfig{file=mavillez-fig8ab.ps,angle=-90,width=4.5in,clip=}
\psfig{file=mavillez-fig8cd.ps,angle=-90,width=4.5in,clip=}
\caption{Evolution of a chimney formed in an OB association located at y=30 pc. The gray scale maps were taken at: (a) 201 Myr, (b) 202 Myr, (c) 203 Myr, (d) 204 Myr after start of calculations.}
\end{figure*} 

The midplane is populated with large regions of cool gas surrounding bubbles of hot gas. The bubbles have different sizes and most of them are isolated in the disk. However, some merge with others forming networks of hot gas. 

As the supernovae occur they change the local structure of the interstellar medium, but are unable to change the global structure. Individual and clustered supernovae dominate the local environment depending on their spatial location. Isolated supernovae change the structure of the inner parts of the disk whereas supernovae in OB associations dominate the upper regions of the disk. During their explosions correlated supernovae easily disrupt the interstellar medium and push material into the halo in collimated structures surrounded by walls of cold gas resembling ``chimneys''. 

Effects like these are observed in Figures 8 and 9 showing the sequential evolution of the disk gas in a region where a chimney evolves. Figure 8 (c) shows the presence of a supernova with a radius of some 50 pc (the supernova is well inside phase III of its evolution) located at $x=400$ pc, $z=-100$ pc. As the shell breaks, the inner parts of the remnant expand into the surrounding medium. A second supernova occurs one million years later at the base of the chimney at $z=-100$ pc (Figure 8 (c)), and in consequence a large amount of hot gas is released into the halo through a tunnel with a width of 110 pc (Figure 8 (d)).

As the supernovae occur in the disk they warm it up to temperatures of $10^{6}$ K. After $5\times 10^{5}$ years, each supernova releases $295 \msolar$ of hot gas (Avillez 1998), contributing to the formation of ``reservoirs'' of hot gas in the disk having enough energy to expand upwards. In its ascending motion the hot gas interacts with the denser gas distributed in the thick disk. Such a configuration is Rayleigh-Taylor unstable and as a result, the hot gas expands with finger-like structures appearing to carve the cooler layers that compose the thick disk (Figure 9). 

In general the disk gas has temperatures varying between $10^{4}$ and $10^{5}$ K filling $\sim 60\%$ of the disk volume. This gas is distributed in the warm ionized layer intermixed with most of the warm neutral gas. Gas with temperatures larger than $10^{5}$ K fills on average a $10\%$ of the disk volume, being the major fraction of such a hot gas located in the thin disk, although a small fraction of it is found in the warm ionized medium. Gas with temperatures smaller than $10^{4}$ K is mainly located at $|z|\leq 500$ pc (filling  on average $25\%$ of the disk volume). 
\begin{figure}[tbhp]
\centering{\hspace*{-0.1cm}}
\psfig{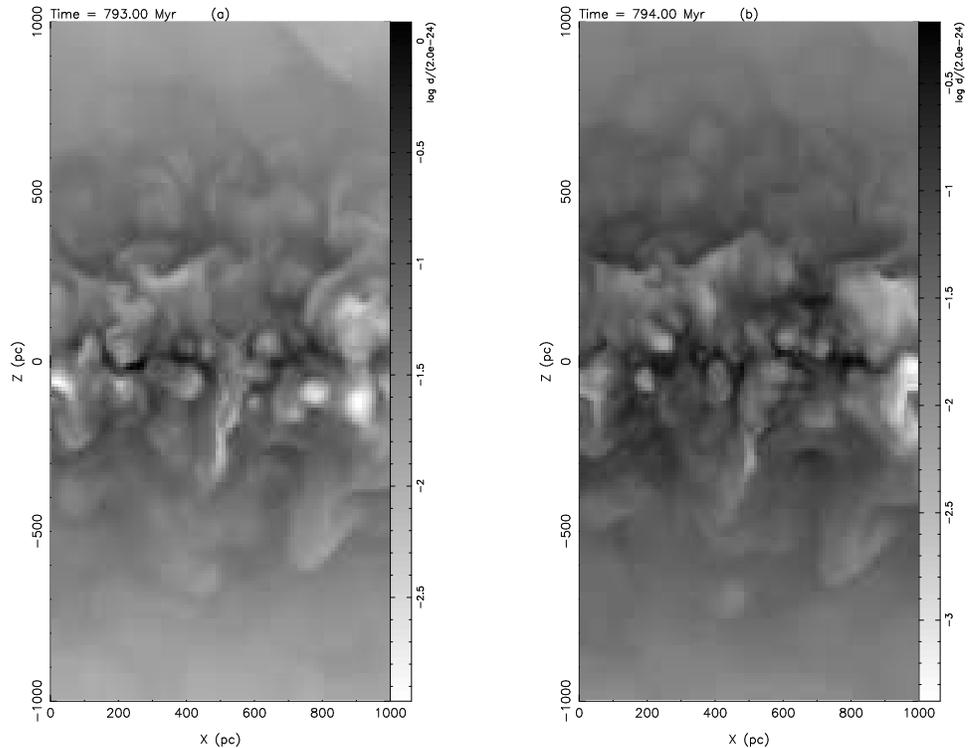}
\caption{Grey-scale maps showing the structure of the Galactic disk measured at $y=980$ pc. The times at which these images refer are: (a) 793 Myr and (b) 794 Myr. The images show that globally the disk has similar structures, although locally there are variations. Note the global expansion of hot gas carving the thick disk and the constant presence of the wiggly cold disk at $z=0$ pc. Local outbursts such as the chimney located at $x=500$, $z=-100$ pc are also observed.}
\end{figure}
Some of this gas is found in the wiggly cool layer located in the midplane as well in the form of small clouds, but the major fraction of it is distributed in the warm neutral layer on either side of the thin disk (see Avillez 1998 for a detailed description).

The warm neutral medium confines the disk gas, preventing it from escaping buoyantly into the halo, unless some highly energetic event occurs in the disk such as correlated supernovae, feeding directly the ionized layers located above the warm neutral medium. The ionized gas extends upwards with a density that decreases smoothly up to 1.4 kpc, showing a steep decrease between this height and 1.5 kpc (Figure 10). At greater heights the gas has lower densities with temperatures greater than $10^{6}$ K. 
The diffuse ionized region located at $1.4\leq z\leq 1.5$ kpc acts as a interface between the thick disk and the halo gas.
\begin{figure}[thbp]
\centering{\hspace*{-0.1cm}}
\mbox{\epsfxsize=2.7in\epsfysize=2.3in\epsfbox{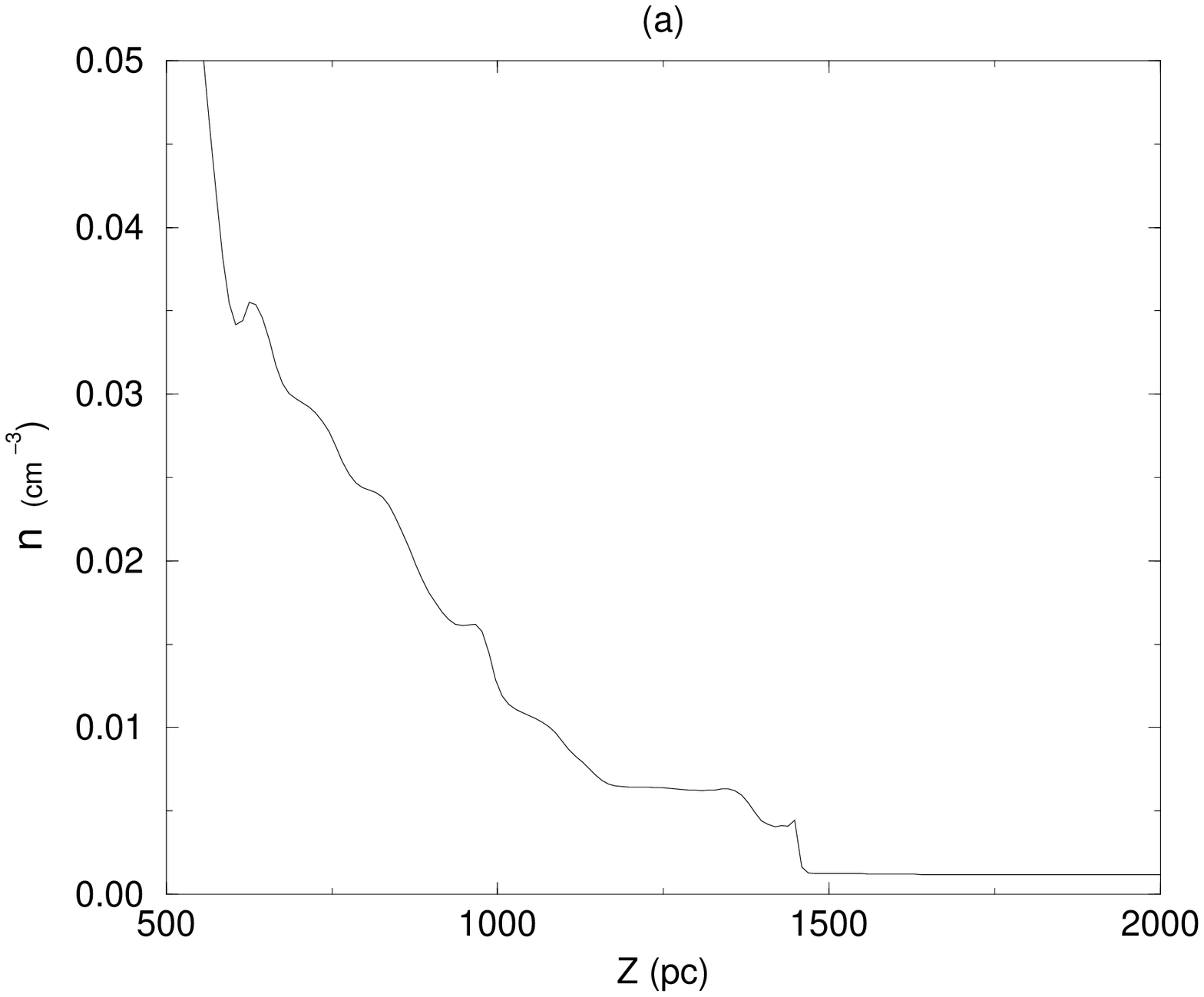}}\mbox{\epsfxsize=2.7in\epsfysize=2.3in\epsfbox{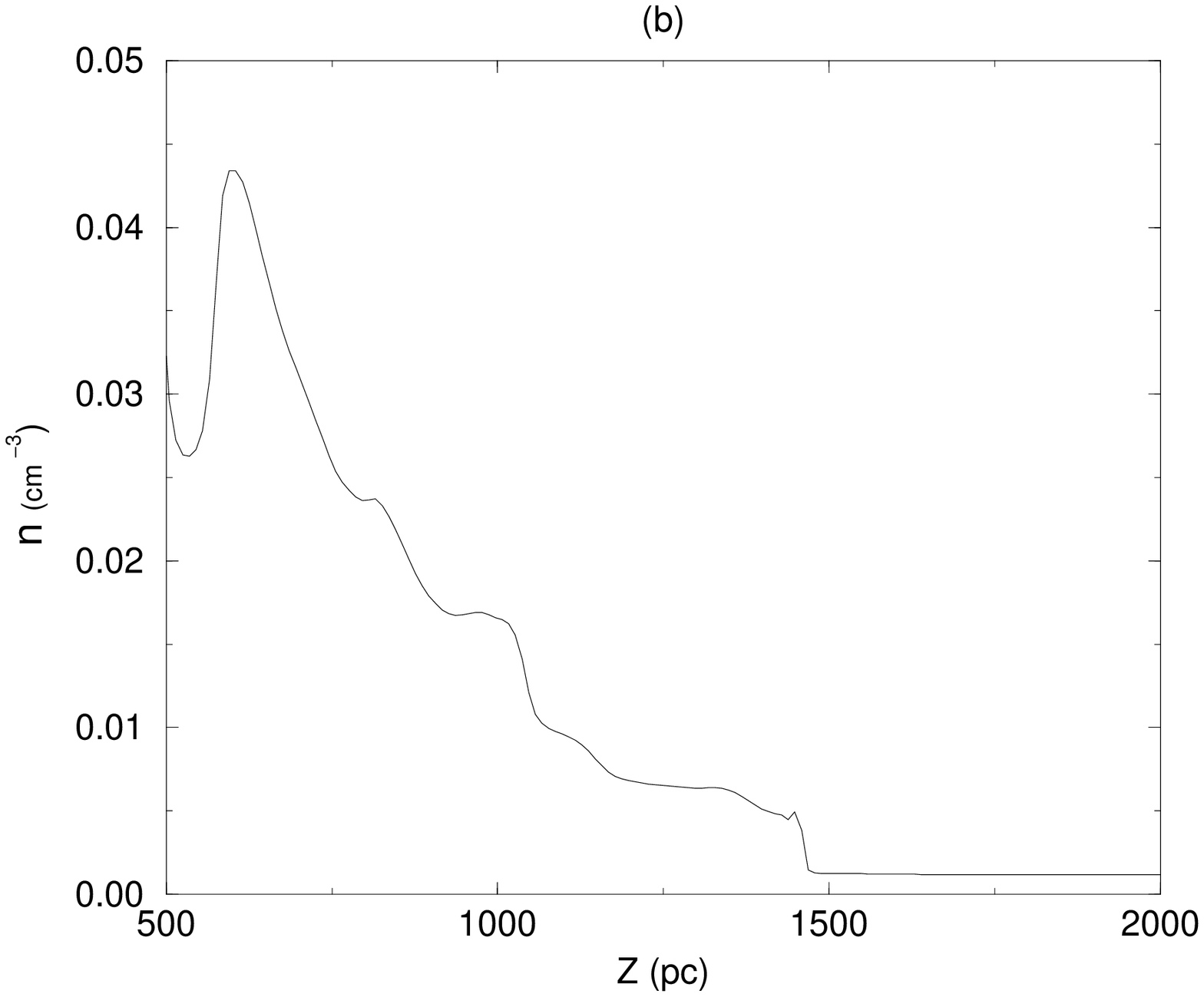}}
\caption{Density profiles of the halo gas measured along two lines of sight passing through the plane at (a) $x=300,\, y=250$ pc and (b) $x=750,\,y=750$ pc. Note the steep variation at $z=1.5$ kpc for both profiles. This shows a clear separation between the diffuse ionized medium know as the Reynolds layer and the halo gas. Thus this medium acts as a intermediate region between the disk and halo.}
\end{figure}
\section{Halo Clouds}
As the hot gas rises into the halo it cools and condenses into clouds of variable sizes that rain back into the Galactic disk. Their sizes vary between a few parsec to tens and hundred of parsec. The clouds are classified, according to their sizes in cloudlets, clouds and complexes, and according to their velocity, $v_{z}$, in low, intermediate, high, and very high velocity clouds. Cloudlets have sizes varying between 5 and 50 pc, whereas clouds have greater sizes. 

Complexes are formed by clouds found close together which have comparable velocities. Their sizes can reach hundreds of parsec. A complex is identified based on the velocity dispersion of their components relative to the bulk of the complex. If the clouds have a velocity dispersion smaller than the local sound speed then they belong to the complex. 

The clouds form as result of small variations in the local density and in places where shock waves intersect. When the gas in the halo is swept up by a shock wave its density is increased by factors of four leading to an increase of its rate of cooling. This process is more efficient if several shocks intersect at the same location in the halo. Figure 11 shows a region in the halo where gas after being swept up by several shock waves starts condensing into clouds - the region located at $200\leq x\leq 400$ pc and $2300\leq z\leq 2600$ pc. 
\begin{figure}[thbp]
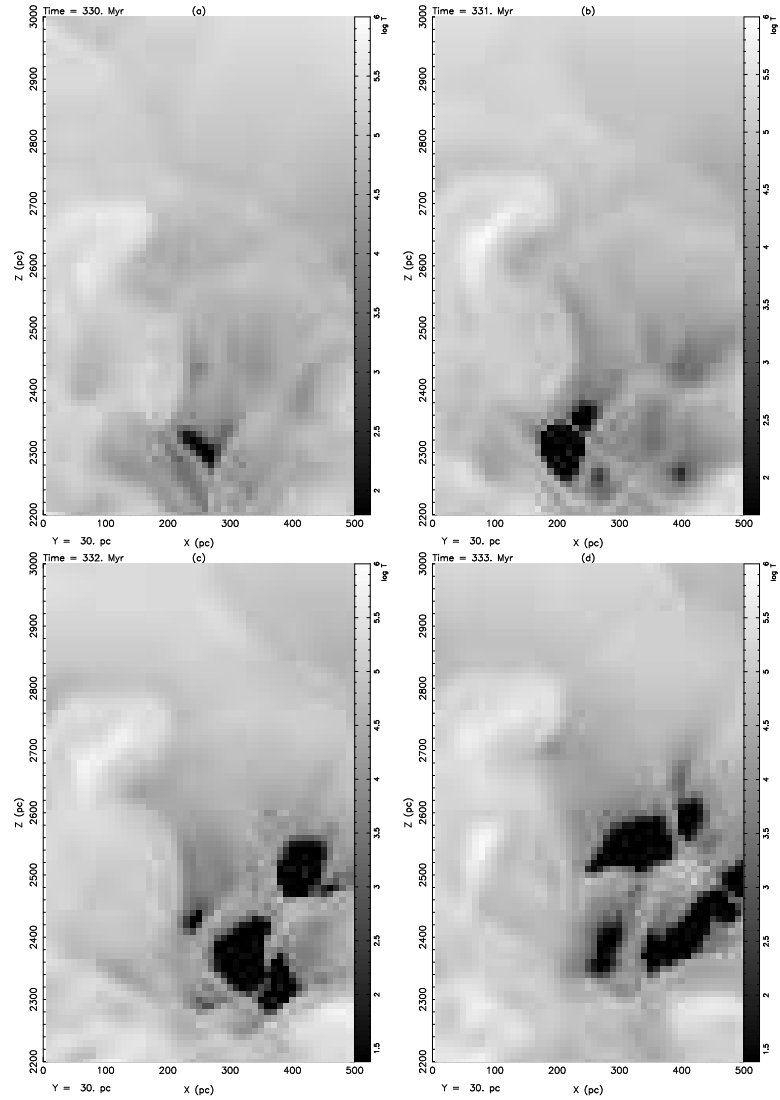

\centering{\hspace*{-0.7cm}}
\psfig{file=mavillez-fig11ab.ps,angle=-90,width=4in,clip=}
\centering{\hspace*{-0.7cm}}
\psfig{file=mavillez-fig11cd.ps,angle=-90,width=4in,clip=}
\caption{Temperature maps of the halo showing the evolution of HI clouds at $2200 \leq z \leq 3000$ pc: (a) 330 Myr, (b) 331 Myr, (c) 332 Myr and (d) 333 Myr. The clouds form as result of the interaction with the hot stream coming from below. The cool gas is swept up by several shock waves, increasing the local density and therefore the cooling of the gas there.}
\end{figure}
The cooler gas is sustained by the hot stream coming from below in an unstable configuration favoring the growth of Rayleigh-Taylor instabilities at the interface separating the two gases, and therefore leading to the formation of the clouds. The clouds surfer further instabilities breaking up into cloudlets that rain over the disk.

Face-on maps of the temperature distribution measured at  $z=2730$ and $z=-1670$ pc (Figure 12) show the effects of a set of clouds in the region they are passing through. 
The clouds interact with hot gas present in that region at the time they arrived there, as well with hot gas moving upwards engulfing them at its passage.

In general the clouds have a sheet-like structure compatible with them being the result of Rayleigh-Taylor instabilities at the base of other clouds located at higher heights.
\begin{figure}[thbp]
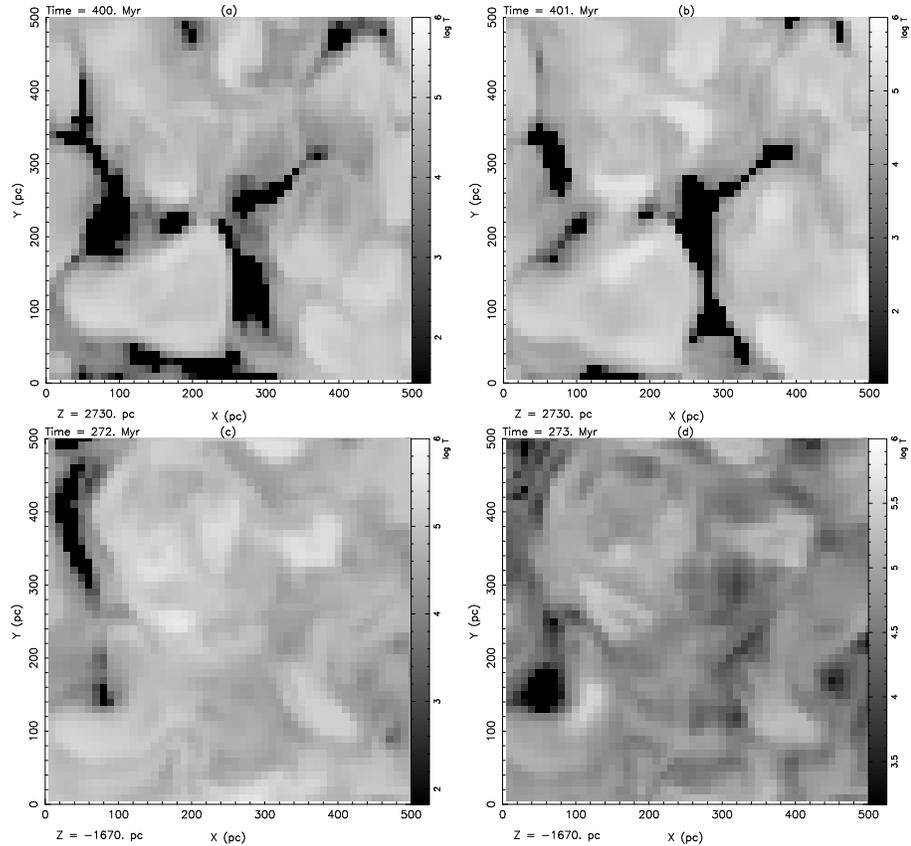

\centering{\hspace*{-0.1cm}}
\psfig{file=mavillez-fig12ab.ps,angle=-90,width=4.7in,clip=}
\psfig{file=mavillez-fig12cd.ps,angle=-90,width=4.7in,clip=}
\caption{Temperature maps showing the evolution of HI clouds at $z=2730$ pc ((a) and (b)) and at $z=-1670$ pc ((c) and (d)). The times these images refer are  (a) 400 Myr, (b) 401 Myr, (c) 272 Myr and (d) 273 Myr. The clouds are passing through the layers and have a multiphase structure with inner cores of cold gas surrounded by warm gas with $T\sim 10^{4}$ K.}
\end{figure}
The clouds have a multiphase structure, where the core of the cloud is formed by cold gas ($T\leq 10^{3}$ K) and surrounded by warmer gas with temperatures of $10^{4}$ K. The clouds are embedded in a hot medium with temperatures greater or equal to $10^{5}$ K (Figures 11 and 12). 

The phases within the cloud have velocity dispersions varying between 1 km s$^{-1}$ to 10 km s$^{-1}$ as can be seen in Figure 13 (a) and (c). 

The figure presents a set of grey-scale images showing the velocity distribution of clouds detected at $z=860$ pc, $z=1200$ pc and $z=-1000$ pc. There is a maximum velocity dispersion of 4 to 5 km s$^{-1}$, having the center of the cloud the smallest z-velocity  and an increase occurs towards the edge. 

The images show the passage of several clouds through the layer they were detected (Figure 13 (c) and (d)). The descending velocity varies from cloud to cloud, where the smaller ones have a larger descending velocity then the larger cloud located at ($x=800$, $y=300$) pc. However, the velocity dispersion between the different clouds is very small. This indicates that they form a complex and are connected by velocity bridges.
\begin{figure}[tbhp]
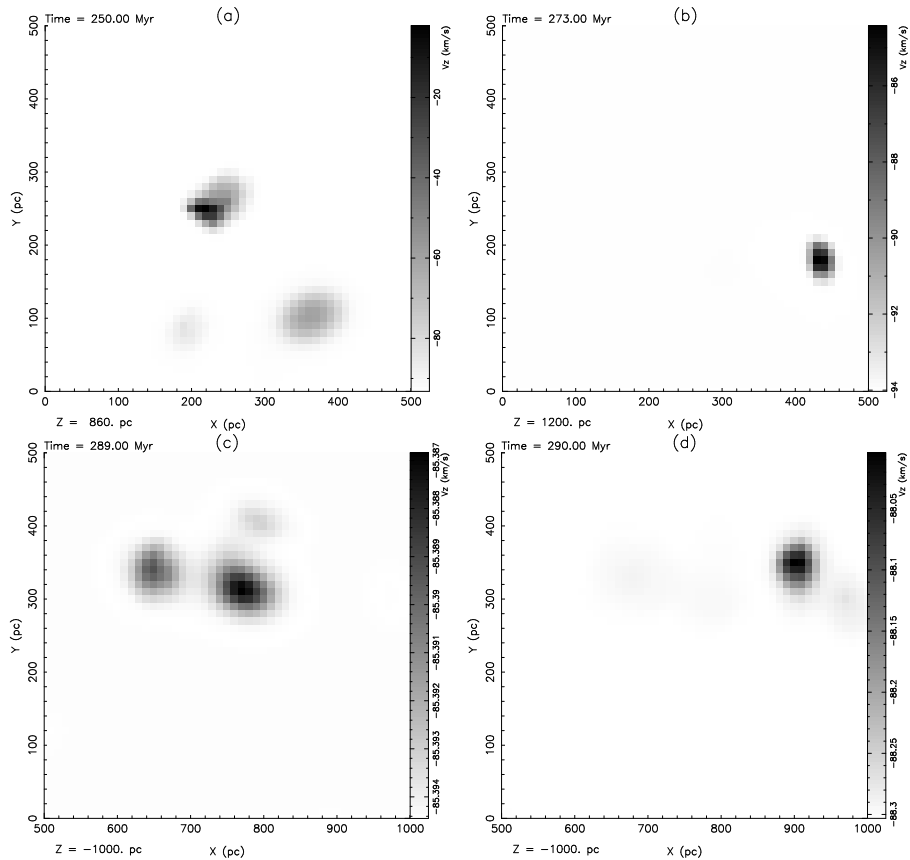

\centering{\hspace*{-0.1cm}}
\psfig{file=mavillez-fig13ab.ps,angle=-90,width=4.7in,clip=}
\psfig{file=mavillez-fig13cd.ps,angle=-90,width=4.7in,clip=}
\caption[Velocity distribution in HI clouds]{Velocity distribution in HI clouds at: (a) 251 Myr and $z=860$ pc, (b) 273 Myr and $z=1200$ pc, (c) 289 Myr and $z=-1000$ pc, (d) 290 Myr and $z=-1000$ pc.}
\end{figure}
The distribution of clouds in the halo vary with $z$. There is no preferable region in the halo where the HI clouds form. However, a large amount of clouds has been detected at heights between 300 pc and 1.5 kpc, with the major distribution of clouds occurring at $z>800$ pc. 

The bulk of the HI clouds has intermediate velocities ($-20$ km s$^{-1}$ up to $-90$ km s$^{-1}$), as can be seen in the histogram presented in Figure 14. $30 \%$ of the clouds have velocities varying between -20 and -40 km s$^{-1}$. 

Only a small fraction of the clouds ($5\%$) has high velocities varying between -90 and -139 km s$^{-1}$. The detection of such a small number of HVCs may be the result of the reduced vertical extent of the grid in this study $\left|z\right|$/[kpc]$\leq 4$, instead of using the full extent up to 10 kpc.

\begin{figure}[thbp]
\centering{\hspace*{-0.1cm}}
\psfig{file=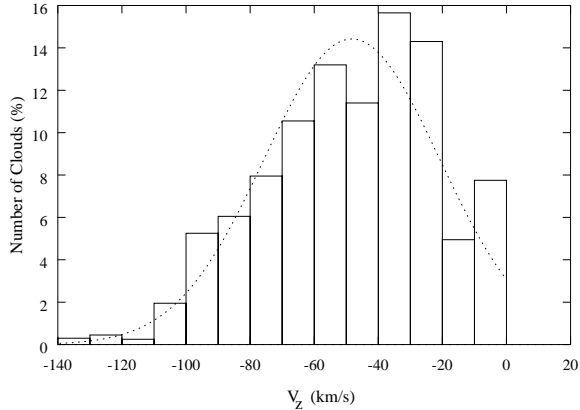,angle=0,width=3in,clip=}
\caption{Distribution of gas with $T<8\times 10^{3}$ K versus the z-velocity during the period of time between 800 Myr and 1 Gyr. The histogram is overlayed by a gaussian with $\sigma=27.4$ km s$^{-1}$ and first quartile=-67.0 km s$^{-1}$.}
\end{figure} 

\section{Discussion and Comparison with Observations}
The striking feature of the simulations is the constant presence of a thin cold wiggly disk gas overlayed by a thick disk of warm gas. The thick disk has two components: a warm neutral layer with a scale height of 500 pc (warm HI disk) and a warm ionized component extending to an height of 1 to 1.5 kpc above the thin HI disk (Table 2). 
\begin{table}[thbp]
\begin{center}
\begin{tabular}{lccc}
\hline
Component & $n$ & $T$ & h$_{z}$\\
& (cm$^{-3}$) &  (K) & (kpc)\\
\hline 
Thin HI disk & $0.6-1$ & $\leq 8\times 10^{3}$ & 0.12 \\
Warm HI disk & $0.1$ & $10^{4}$ & 0.5 \\
Warm HII disk & $0.01-0.02$ & $10^{4} - 10^{5}$ & 1 - 1.5\\
\hline
\end{tabular}
\caption{Physical properties of the thin and thick disk gas.}
\end{center}
\end{table}

The vertical distribution of the layers that compose the Galactic disk are compatible with observations carried out by Lockman (1984), Lockman {\rm et~al.} (1986) and Reynolds (1987) who found that the disk has warm neutral and ionized components. Lockman identified a thin HI disk with a gaussian distribution with $\sigma_{z}=135$ pc and an exponential distribution with a scale height of some 500 pc. Reynolds observed a warm ionized medium extending far beyond the neutral layer with scale heights of 1 kpc.

The simulations show both layers are fed with the hot gas coming from the thin disk through two major processes: large scale outflows and chimneys. Chimneys result from correlated supernovae generating superbubbles which blow holes in the disk, provided the superbubbles are displaced at some 100 pc from the Galactic plane and located in the vicinity of a region with local density gradients in the $z$-direction(Figures 8 and 9). 

The supershells are accelerated in the vertical direction and high pressure gas forces its way out through relatively narrow channels (chimneys) with widths of some 100 - 150 pc.

Not all superbubbles create chimneys. In those cases, the superbubbles as well as isolated supernovae expand and die within the disk, releasing large amounts of hot gas with enough energy to be held gravitationally expanding buoyantly upwards to generate large scale outflows (Figure 9). The ascending flow triggers the growth of Rayleigh-Taylor instabilities as it interacts with the cooler denser medium in the thick disk. As a consequence the hot gas acquires finger like-structures with mushroom caps in their tips. As the fingers expand they suffer further instabilities until they cool and merge with the warm gas. 

The gas in the finger cools from outside towards the center of the finger, thus giving it the appearance of being enveloped by a thin sheet of cooler gas (Figures 3, 8 and 9). A structure called the ``anchor'' and having properties similar to those described above has been observed in the southern Galactic hemisphere and reported by Normandeau \& Basu (1998). 

The kinematics and morphology of the bubbles in the disk are similar to that observed in face-on galaxies M31 (Brinks \& Bajaja 1986) and M33 (Deul \& Hartog 1988). These surveys revealed the presence of roughly elliptical features with a relative absence of neutral gas and having sizes varying between 40 and 150 pc. Some of these regions show a clear shell structure, indicating a relationship to isolated or clustered supernovae (Deul \& Hartog 1988).

As the hot gas ascends into the halo it cools condensing into clouds that rain back over the disk. Their distribution in the halo varies with $z$. The major fraction of clouds have intermediate velocities and are mainly found between 800 and 2.2 kpc.

The clouds have a multiphase structure with the cold gas having temperatures of some $10^{3}$ K embedded in a warmer phase. A local analysis the clouds shows components of different velocities with dispersions up to 20 km s$^{-1}$. The clouds found closed together have smaller velocity dispersions between them, suggesting the presence of velocity bridges connecting the clouds.

These results are compatible with observations regarding to the location and distribution of IVCs in the halo (Wesselius \& Fejes (1973), Kuntz \& Danly (1996)) as well with theoretical predictions of Houck \& Bregman (1990) in placing the major fraction of these clouds between 1 and 2 kpc, and observations of the clouds internal structure (Cram \& Giovanelli 1976; Shaw {\rm et~al.} 1996; see Wolfire {\rm et~al.} 1995 for an interpretation of these observations). 

Only a small number of high velocity clouds have been detected in the simulations. This suggests that most of the high velocity clouds are formed at heights greater than 4 kpc, and therefore would not be detectable in the simulations. Their formation within the Galactic fountain is only possible provided the fountain gas rises to some 10 kpc. Such a scale height is incompatible with the analytical models described in \S1, unless the injection level of the fountain is shifted some 1.5 kpc above the plane.

\section{Conclusions: Where Starts the Galactic Fountain?} 
The Galactic fountain is defined as the cycle performed by the gas escaping from the Galactic disk and rising into the halo, where eventually it cools and condenses into clouds. Such a fountain includes the outflows from isolated as well as clustered supernovae scattered in the Galactic disk generating large scale as well as localized outbursts causing the formation of an unstable mixed structure of neutral and ionized gases spread in the disk and halo.

The upper parts of the thick ionized disk, located at some 1.4 kpc above the plane, act as a disk-halo interface where the hot plasma expands upwards in a smooth flow. This process resembles the classical fountain predicted by Shapiro \& Field (1976), Bregman (1980) and Kahn (1981) but with the injection level located  at some 1.5 kpc above the Galactic plane. 

The theoretical descriptions of Kahn (1981) can then be applied to this flow. The flow starts subsonic, becomes supersonic at some 3.4 kpc above the injection level, cools and continues its motion ballistically for more 4.4 kpc (Avillez 1998) reaching the maximum height at $z\sim 9.3\pm 1$ kpc. The clouds that will be formed during the descent of the cold gas will have, at some height, high velocities (a detailed study is presented in Avillez 1999). 

The descending clouds headed by a shock, sweep up the halo gas in their passage, thereby shocking and heating it. A thin layer of shocked gas is formed between the shock and the descending cloud. The configuration is Rayleigh-Taylor unstable and the cloud breaks up into cloudlets with sheet-like forms. These structures rain over the disk with intermediate velocities (Berry {\rm et~al.} 1997; Avillez 1998) at an inflow rate of 2.97 $\msolar$ yr$^{-1}$ on either side of the Galactic plane.

\acknowledgments 
This paper is dedicated to the memory of Prof. Franz D. Kahn. He will be 
sadly missed. I would like to thank Ana Inez Gomes de Castro, Bart Wakker, 
Brad Gibson and 
Joel Bregman for their careful reading of the manuscript and suggestions which 
lead to its improvement.

\end{document}